\documentclass[fleqn,usenatbib]{mnras}

\usepackage{newtxtext,newtxmath}
\usepackage[T1]{fontenc}

\usepackage{tablefootnote}
\usepackage{soul}
\DeclareRobustCommand{\VAN}[3]{#2}
\let\VANthebibliography\thebibliography
\def\thebibliography{\DeclareRobustCommand{\VAN}[3]{##3}\VANthebibliography}

\usepackage{graphicx}	% Including figure files
\usepackage{amsmath}	% Advanced maths commands

\usepackage{xcolor}

\title[uGMRT study of low-mass clusters]{uGMRT detection of cluster radio emission in low-mass Planck~SZ clusters}

\author[Paul S. et al.]{Surajit Paul,$^{1}$\thanks{E-mail: surajit@physics.unipune.ac.in}
Prateek Gupta,$^{1}$
Sameer Salunkhe,$^{1}$
Shubham Bhagat,$^{1}$
Satish Sonkamble,$^{2}$\newauthor
Manish Hiray,$^{1}$
Pratik Dabhade,$^{3,4,5}$ and Somak Raychaudhury$^{4,6}$
\\\\
$^{1}$ Department of Physics, Savitribai Phule Pune University, Pune 411007, India\\
$^{2}$ INAF-Padova Astronomical Observatory, Vicolo dell’Osservatorio 5, I-35122 Padova, Italy \\
$^{3}$ Leiden Observatory, Leiden University, Niels Bohrweg 2, 2333 CA, Leiden, Netherlands. \\
$^{4}$ Inter-University Centre for Astronomy and Astrophysics, Pune 411007, India\\
$^{5}$ Observatoire de Paris, LERMA, Coll\`ege de France, CNRS, PSL University, Sorbonne University, 75014, Paris, France \\
$^{6}$ School of Physics and Astronomy, University of Birmingham,Birmingham B15~2TT, UK\\
\vspace{-0.9cm}}

\date{Accepted XXX. Received YYY; in original form ZZZ}
\pubyear{2015}

\begin{document}
\label{firstpage}
\pagerange{\pageref{firstpage}--\pageref{lastpage}}

\maketitle


\begin{abstract}
Low-mass ($M_{\rm{500}}<5\times10^{14}{\rm{M_\odot}}$) galaxy clusters have been largely unexplored in radio observations, due to the inadequate sensitivity of existing telescopes. However, the upgraded GMRT (uGMRT) and the Low Frequency ARray (LoFAR), with unprecedented sensitivity at low frequencies, have paved the way to closely study less massive clusters than before. We have started the first large-scale programme to systematically search for diffuse radio emission from low-mass galaxy clusters, chosen from the Planck Sunyaev-Zel'dovich cluster catalogue. We report here the detection of diffuse radio emission from four of the 12 objects in our sample, shortlisted from the inspection of the LoFAR Two Meter Sky Survey (LoTSS-I), followed up by uGMRT Band 3 deep observations. The clusters PSZ2~G089 (Abell~1904) and PSZ2~G111 (Abell~1697) are detected with relic-like emission, while PSZ2~G106 is found to have an intermediate radio halo and PSZ2~G080 (Abell~2018) seems to be a halo-relic system. PSZ2~G089 and PSZ2~G080 are among the lowest-mass clusters discovered with a radio relic and a halo-relic system, respectively. A high ($\sim30\%$) detection rate, with powerful radio emission ($P_{1.4\ {\rm GHz}}\sim10^{23}~{\rm{W~Hz^{-1}}}$) found in most of these objects, opens up prospects of studying radio emission in galaxy clusters over a wider mass range, to much lower-mass systems.
\end{abstract}

\begin{keywords}
 Galaxies: clusters: general -- Radio continuum: general -- X-rays: galaxies: clusters -- large-scale structure of Universe
 \vspace{-0.9cm}
\end{keywords}
%%%%%%%%%%%%%%%%%%%%%%%%%%%%%%%%%%%%%%%%%%%%%%%%%%

%%%%%%%%%%%%%%%%% BODY OF PAPER %%%%%%%%%%%%%%%%%%

\section{Introduction}\label{intro}

Observations of diffuse radio emission due to synchrotron processes reveal the dynamical state and the non-thermal energy evolution in galaxy clusters. Though massive clusters ($>5\times10^{14}{\rm{M_\odot}}$) have been detected in abundance at radio-frequencies over the last few decades \citep[e.g.][]{Weeren_2019SSRv}, the less massive ones ($<5\times10^{14}{\rm{M_\odot}}$) have remained largely unexplored. This is mainly due to limitations in sensitivity at the usual radio-frequencies. Further, the dearth of theoretical estimates for favourable detections in low-mass systems, with the currently available instruments, have so far discouraged systematic studies. For instance, while the theoretical prediction of \citet{Cassano_2010A&A} shows a low probability for halo detection in low-mass clusters with the Low Frequency ARray (LoFAR), the recent analysis of VLA and GMRT data of 75 high mass clusters with radio haloes by \citet{Cuciti_2021A&A} shows a large scatter in the radio power-mass correlation, leading to  broader confidence levels at the lower-mass end, indicating the prospect of a better detectable fraction of radio haloes from them. It can be noted that, before uGMRT and LoFAR, only a handful of low-mass clusters had been detected with cluster radio emission \citep{Bernardi_2016MNRAS,Gasperin_2017A&A,Kale_2017MNRAS,Dwarakanath_2018MNRAS,Botteon_2019A&A,Knowles_2019MNRAS}. However, in the absence of their detection, the understanding of the hierarchical formation of clusters, especially in the non-thermal regime, would remain incomplete.

Radio emission linked to the tenuous plasma of the intracluster medium (ICM) is diffuse in nature, with very low surface brightness ($\sim\mu\rm{Jy~arcsec^{-2}}$) and steep spectrum ($\alpha < -1$, with $S_\nu \propto \nu^{\alpha}$). Such detections are usually hundreds of kpc to Mpc in extent, and are mostly found in merging systems. Depending on their location in the cluster and the possible process of origin, they are divided into two major classes \citep{Feretti_2012A&ARv}. Such diffuse and feebly ($\sim5\%$) polarised centrally-located, emission, possibly resulting from the turbulence in the ICM, is generally termed a `radio halo'. In contrast, comparatively strongly polarised ($\sim30\%$) `radio relics' usually originate from cluster merger shocks, and are mostly located at the cluster outskirts. Objects of a third kind, with central diffuse radio emission, but smaller in extent ($100-500$~kpc), are known as radio `mini-haloes'. These are usually found around the brightest cluster galaxy (BCG) in relaxed clusters with cool-cores, and their emission is thought to be of hadronic origin, i.e., from the secondary electrons produced from the central AGN \citep[e.g.][]{Ferrari_2008SSRv,Giacintucci_2019ApJ,Ignesti_2020A&A}.

Diffuse radio sources in clusters have so far been detected mostly in massive systems \citep{Weeren_2019SSRv}. However, in various studies of low-mass clusters, evidence of mergers and non-gravitational processes \citep[e.g.][]{Lovisari_2015A&A} is frequently observed, and AGN feedback processes are known to significantly affect such systems \citep[e.g.][]{Gaspari_2011MNRAS}. Recent simulations of the evolution of turbulence and cosmic rays in low-mass systems show significant deviations from scaling relations found in richer clusters, having flatter slopes and higher fluctuations \citep[e.g.][]{Paul_2017MNRAS,John_2019MNRAS}, indicating that more of them should be detectable because of their non-thermal properties. Furthermore, as low-mass systems are cooler than the massive ones, the Mach number of the shocks in the ICM, which depends on the ICM temperature, would be noticeably higher in these sources \citep{Sarazin_2002ASSL}. This would mean more and more low-mass clusters would become observable with diffuse radio emission as the sensitivity of radio telescopes improves, as is already happening (see, e.g., review by \citealt{Weeren_2019SSRv}). The unprecedented sensitivity of LoFAR ($\sim100~\mu$Jy at 144~MHz) has revolutionised the study of diffuse emission in low-mass systems \citep{Hoang_2019A&A,Knowles_2019MNRAS,Botteon_2019A&A,Paul_2020A&A}, particularly the emission linked with the ICM in groups \citep{Nikiel_2019A&A}.

Our study begins with a serendipitous discovery of a radio relic and trailing emission in a low-mass cluster, Abell~1697 \citep{Paul_2020A&A}, from the LoTSS data release~1 (LoTSS-I). Thereafter, we started a programme to systematically search for low-mass ($M<5\times10^{14}{\rm{M}}_\odot$) systems listed in the Planck~SZ galaxy cluster catalogue (\citealt{Planck_2016A&A}; Planck-16 hereafter), in the region overlapping that of LoTSS-I. Mosaic images from LoTSS-I are reported to suffer from different types of artefacts, such as low-level positive or negative haloes  \citep{Shimwell_2019A&A}, making it difficult to precisely characterise faint diffuse sources, particularly since the raw data are not publicly available for reprocessing. This led us to obtain fresh deep observations at a complementary frequency, with uGMRT at Band~3, to confirm the Planck-SZ systems for which we found evidence of diffuse emission in LoTSS-I. Further, to understand the nature of the detected radio emission, we went on to study the optical and X-ray properties of these systems from publicly-available archival data. 

In this paper, we report the detection of cluster radio emission from four low-mass galaxy clusters. Following the details of data selection and analysis presented in \S~\ref{sec:data}, the LoTSS-I radio detection, the follow-up uGMRT observations, along with multi-wavelength properties of these systems are discussed in \S~\ref{sec:result}. Our conclusions are outlined in \S~\ref{sec:conc}. A  $\Lambda$CDM cosmology is assumed with parameters $H_{0}$=$70~{\rm{kms^{-1} Mpc^{-1}}}$, $\Omega_M$=$0.3$, $\Omega_{\Lambda}$=$0.7$ throughout this paper.

\begin{table}
\caption{Selected low-mass clusters from the Planck-SZ cluster list and LoTSS-I maps}
\centering
\small
%\arraystretch{1.1}
\begin{tabular}{lcccc}
\hline
Object & $M^{SZ}_{500} M_{\odot}$ & z & LoTSS & RMS noise \\
& ($10^{14}$)& &indication & ($\mu$Jy~B$^{-1}\dagger$)\\
\hline
{\bf *} PSZ2~G080.16+57.65 & 2.51 & 0.08780 & H\&R & 240 \\
PSZ2~G088.98+55.07 & 4.92 & 0.70235 & NDE & 180\\
{\bf *} PSZ2~G089.52+62.34 & 1.83 & 0.07008  & cR & 200\\
PSZ2 G095.22+67.41 & 1.50 & 0.06250 & NDE & 100\\
PSZ2~G096.14+56.24 & 2.77 & 0.13977 & NDE & 160\\
PSZ2 G098.44+56.59 & 2.83 & 0.13184 & NDE & 160\\
PSZ2~G099.48+55.60 & 2.81 & 0.10510  & Outside & --\\
{\bf *} PSZ2 G106.61+66.71 & 4.67  & 0.33140 & iH? & 300\\
{\bf *} PSZ2 G111.75+70.37 & 4.34 & 0.18300 & R\&T & 420\\
PSZ2 G118.34+68.79 & 3.77 & 0.25488 & NDE & 150\\
PSZ2 G123.66+67.25 & 4.38 & 0.28380 & NDE & 130\\
PSZ2 G136.92+59.46 & 1.81 & 0.06500 & Outside & --\\
PSZ2 G144.33+62.85 & 2.66 & 0.13200 & NDE & 120\\
PSZ2 G145.65+59.30  & 4.73 & 0.34748 & NDE & 110\\
\hline
\end{tabular}\label{tab:sample}\\
\footnotesize{Here
{\bf *} indicates observed with uGMRT; NDE=No Diffuse Emission. Structures are indicated as Relic (R), Halo (H), Intermediate Halo (iH) and Trailing emission (T). $\dagger$The beam size (B) is $20\arcsec\times20\arcsec$ for all the objects except for PSZ2~G111, for which it is $25\arcsec\times25\arcsec$.}
\end{table}

\section{Data selection and analysis}\label{sec:data}

We produced a list of low-mass clusters ($\lesssim 5 \times10^{14}~{\rm{M}}_\odot$), from a total of 1653 objects published in the latest Planck~SZ galaxy cluster catalogue (Planck-16). Since our goal is to first check the  LoTSS-I \citep{Shimwell_2019A&A} for evidence of diffuse emission, before following-up with uGMRT, we select only those located in the area of the sky between RA $161^\circ .25$ to $232^\circ .5$  and DEC $+45^\circ$ to $+57^\circ$.
This resulted in 14 objects (see Tab.~\ref{tab:sample}), for which the LoTSS images are publicly available. We then visually inspected each of them in LoTSS-I maps and found that two of these fall outside the edges of the provided mosaics. Finally, among the remaining 12 clusters, four (PSZ2 G080.16+57.65 (PSZ2 G080), PSZ2 G089.52+62.34 (PSZ2 G089), PSZ2 G106.61+66.71 (PSZ2 G106) and PSZ2 G111.75+70.37 (PSZ2 G111)) were found to have indications of diffuse radio emission, whereas the rest of the eight objects are devoid of any diffuse emission at the local RMS noise level as given in Tab.~\ref{tab:sample}. Of these, images of PSZ2~G096.14+56.24 and PSZ2~G123.66+67.25 are significantly affected by the PSF side-lobes of central bright sources (flux $>1$~Jy). Independently, all four of these appear in the LoTSS sample of cluster radio sources  \citep{Weeren_2020arXiv}. We then obtained the uGMRT \citep{Gupta_2017CSci} Band~3 (300-500 MHz) data (proposal code: ddtC~157) with $\sim2$~hours on-source time on each of these objects.

\begin{table}
\caption{uGMRT imaging parameters. Those for the point source subtracted images are given within brackets.}
\centering
\small
\begin{tabular}{lcccc}
\hline
Object & robust & taper~(\arcsec) & beam~(\arcsec) & RMS ($\mu$Jy~B$^{-1}$)  \\ 
\hline

PSZ2~G080 & 0.5(1.0) & 10.0(8.0) & $34\times17$($65\times24$) & 55(120)\\

PSZ2~G089 & 0.5 & 5.0 & $15\times12$ & 50 \\
PSZ2~G106 & 0.5(1.0) & 0.0(10.0) & $11\times7$($35\times17$) & 45(120)\\

PSZ2~G111 & 0.0/1.0 & 10.0/10.0 & $13\times10$/$21\times16$ & 65/100\\
\hline
\end{tabular}\label{tab:impara}
%\vspace{-0.3cm}
\end{table}

Data reduction and imaging were primarily performed using the {\sc{SPAM}} pipeline (for details see \citealt{Intema_2017A&A}). This is a powerful data analysis and imaging pipeline that takes care of direction-dependent variations (i.e., due to antenna beam pattern and due to ionosphere) in visibility, amplitude and phase, across the field of view. Using SPAM, wide-band uGMRT data were first split into six narrower bands, and each of them were initially calibrated for flux and band-pass using the source 3C286.  The final SPAM-calibrated data for each narrow-band, after imaging, were combined to produce wide-band images using the WSclean package \citep{Offringa_2014MNRAS}, with the specific parameters as mentioned in Tab.~\ref{tab:impara}. To make point source subtracted low resolution images, high-resolution point-source only images were first produced using an appropriate "uvcut" in WSclean and then the point sources were subtracted from the UV-data, using the FT and UVSUB tasks in CASA. 

The radio flux densities (S) were measured within $3\sigma$ contours of each map ($\sigma$ being the local rms noise), and flux density errors were estimated ($\sigma_S$) using the relation $\sigma_S=\sqrt{(f{\times}S)^2+N\sigma^2+\sigma_{sub}^2}$, where $N$ is the number of beams within $3\sigma$ contours and $\sigma_{sub}$ is the uncertainty due to compact source subtraction (applied to the point source subtracted maps). We assumed an usual $f$=10\% for uGMRT and $f$=20\% for LoTSS-I (as mentioned in \citealt{Shimwell_2019A&A}) of $S$ to be the error due to calibration uncertainties. In addition to the usual method of flux density measurements within $3\sigma$ isophotes in radio images, we employed a more generic method to compute the radio flux density of halo-type diffuse radio sources in galaxy clusters, by fitting azimuthally-averaged brightness profiles with an exponential function \citep{Murgia_2009A&A}. The fitted exponential profile is of the form $I(r) = I_0 \exp{-(r/r_e)}$,
where $r_e$ is the characteristic e-folding radius and $I_0$ is the peak surface brightness of the radio halo. Profiles were fit using a least-squares fitting algorithm. For statistical purposes, in each radio image, the surface brightness was sampled several times with different linear radial step sizes. Corresponding to the fitting parameters, $I_0$ and $r_e$, for each surface brightness sampling, the flux densities were measured by integrating the fitted brightness profile upto three times the e-fold radius, i.e. $3r_e$, following \citet{Murgia_2009A&A}. Accordingly, the statistical mean and standard deviation of fitted flux densities for each halo are reported in Tab.~\ref{tab:findings}. All images used in this paper were manipulated and analysed using the CASA package \citep{McMullin_2007ASPC}.

\begin{table*}
\caption{Radio (uGMRT:~400~MHz \& LoTSS-I:~144~MHz (for the same area as uGMRT observations)) and X-ray (XMM:~0.5-8~keV \& {\it Chandra}:~0.5-3~keV) properties.}
\centering
\small
\begin{tabular}{rlccccccc}
\hline
Object & Abell name & $S_{400} (3\sigma)$ & LoTSS-I flux*& $S_{144}{\dagger}$ & Size (uGMRT) & \multicolumn{2}{c}{Model Halo}& $L_{\rm{X}}$ \\
& & (mJy) & (mJy)& (mJy) & (kpc) & [Flux (mJy) & Diameter (kpc)] & $(10^{44}\rm{erg~s}^{-1})$ \\
% & (mJy) & ${\rm{W~Hz^{-1}}}$) & \rm{erg}~$\rm{s}^{-1}$)  \\ 
\hline

PSZ2~G080~(H) & Abell~2018 & $19.3\pm2.0$ & $72.1\pm17.1$ & $92\pm32$ & $540\times490$ & $18.0\pm0.5$ & $493\pm21$ & $0.81^{+0.01}_{-0.01}$\\
(R)& &$13.3\pm1.4$& -- & $55.8\pm11.6$ & 1000 & -- & -- & --\\
PSZ2~G089~(cR) & Abell~1904 &  $6.6\pm 0.7$ & $19.9\pm 3.1$ & $74.6\pm15.0$ & $205\times45$ & -- & --& $2.54^{+0.03}_{-0.02}$ \\

PSZ2~G106~(iH) & -- & $8.6\pm1.0$ & $23.9\pm4.9$  & $20\pm4$ & $550\times530$ & $6.6\pm0.2$ & $507\pm17$ & $3.97^{+0.40}_{-0.10}$\\
PSZ2~G111~(R) & Abell~1697 & $49.7\pm5.0$ & $130.2\pm26.1$ & $106.7\pm21.4$ & $750\times350$  & -- & -- & $5.26^{+0.30}_{-0.40}$\\
~(T)& & $15.3\pm1.6$ & $41.7\pm8.3$ & -- & $450\times350$& --& -- & -- \\
\hline
\end{tabular}\label{tab:findings}\\
\footnotesize{Here H, R and T represents halo, relic and trailing emission. `*' Flux density for the same area as uGMRT $3\sigma$ area. `$\dagger$' As reported in \citet{Weeren_2020arXiv}.}

\end{table*}

For multi-wavelength follow-up, we obtained PanSTARRS-1 (optical) data for all the objects, and, in X-rays, XMM-Newton data of $\sim$35~ks (ObsID:0821040401), $\sim$15~ks (ObsID:0821040301) and $\sim$44~ks (ObsID:0827020801) for PSZ2~G080, PSZ2~G089 and PSZ2~G111, respectively and the {\it Chandra} observations for PSZ2 G106 (15~ks, ObsID:21708). The {\it Chandra} data were reduced using CIAO~v4.11, CALDB~4.8.4 following the standard data reduction threads and using re-projected blank-sky background files. The XMM-Newton data were analysed using SAS~v18 and the latest calibration files. We used the ``IMAGES" tool to clean high background flares, apply exposure corrections, do spatial smoothing, and create and combine PN and MOS images. Spectral extraction and fitting were performed in XSPEC v12.10.1 with fitted {\tt tbabs$\times$apec} model in 0.5--8.0 keV energy band to compute X-ray luminosities within $r_{500}$ of each sources. The X-ray surface brightness profiles were extracted and fitted using PROFFIT v1.5 \citep{eckert2011}.

\section{Detection from LoTSS-I and {\lowercase{u}}GMRT Band~3}\label{sec:result}

The measured parameters of these newly detected objects with cluster radio emission from uGMRT and LoTSS-I are outlines in Tab.~\ref{tab:findings}. The details of individual objects are individually presented below.

\begin{figure*}
\includegraphics[width=9.0cm]{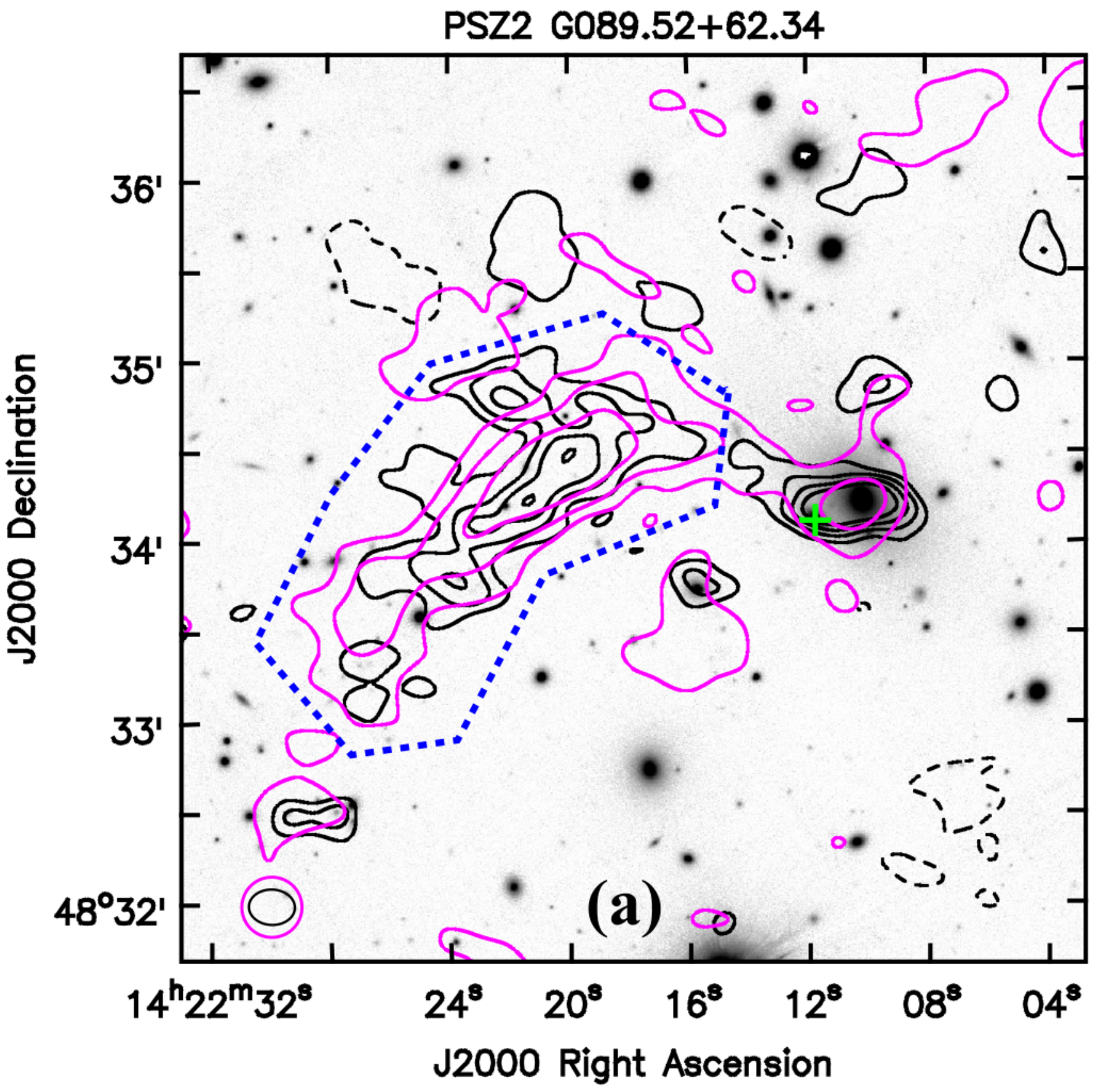}
\includegraphics[width=8.5cm]{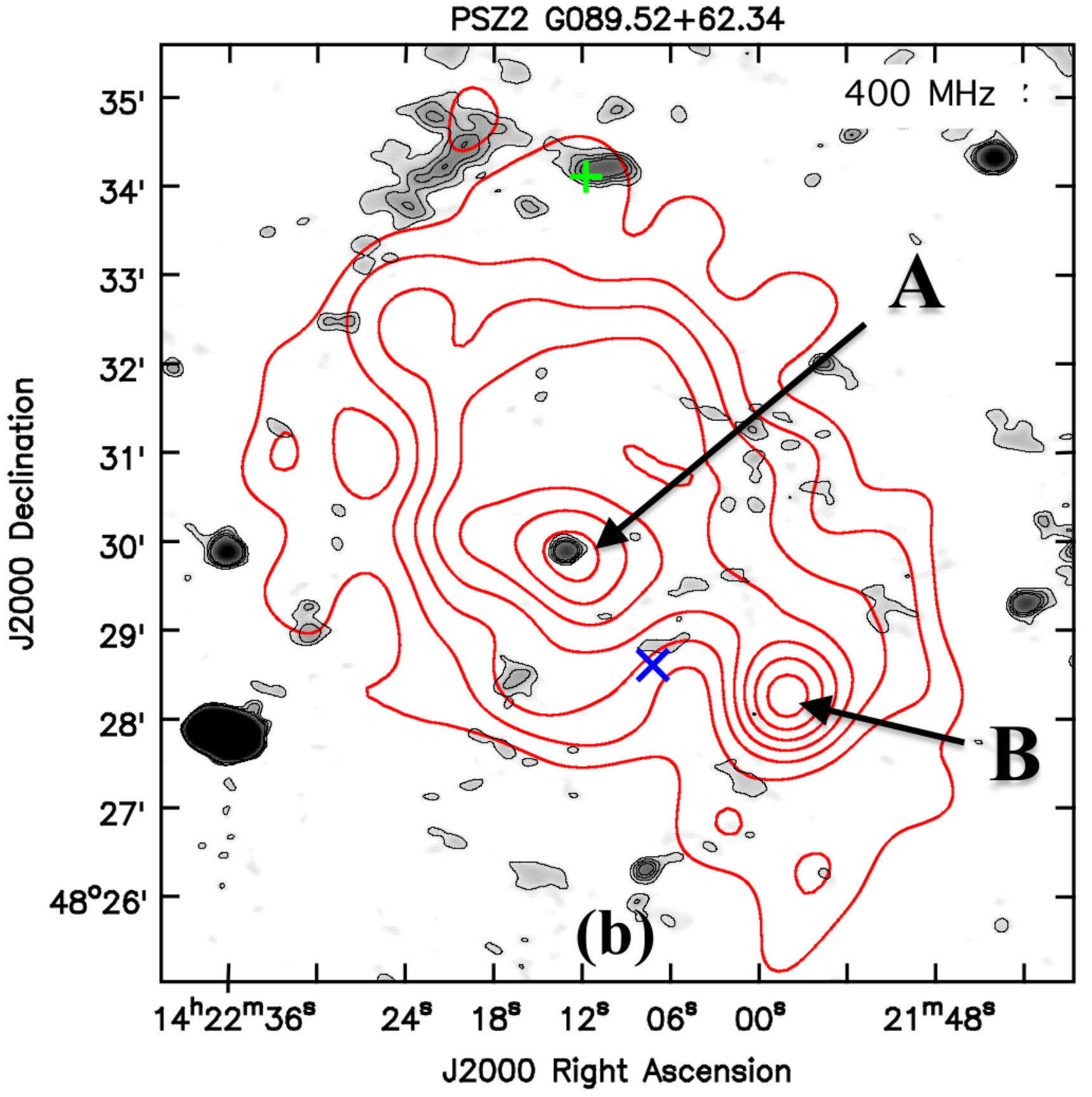}\hspace{0.0cm}\\
\includegraphics[width=9.2cm]{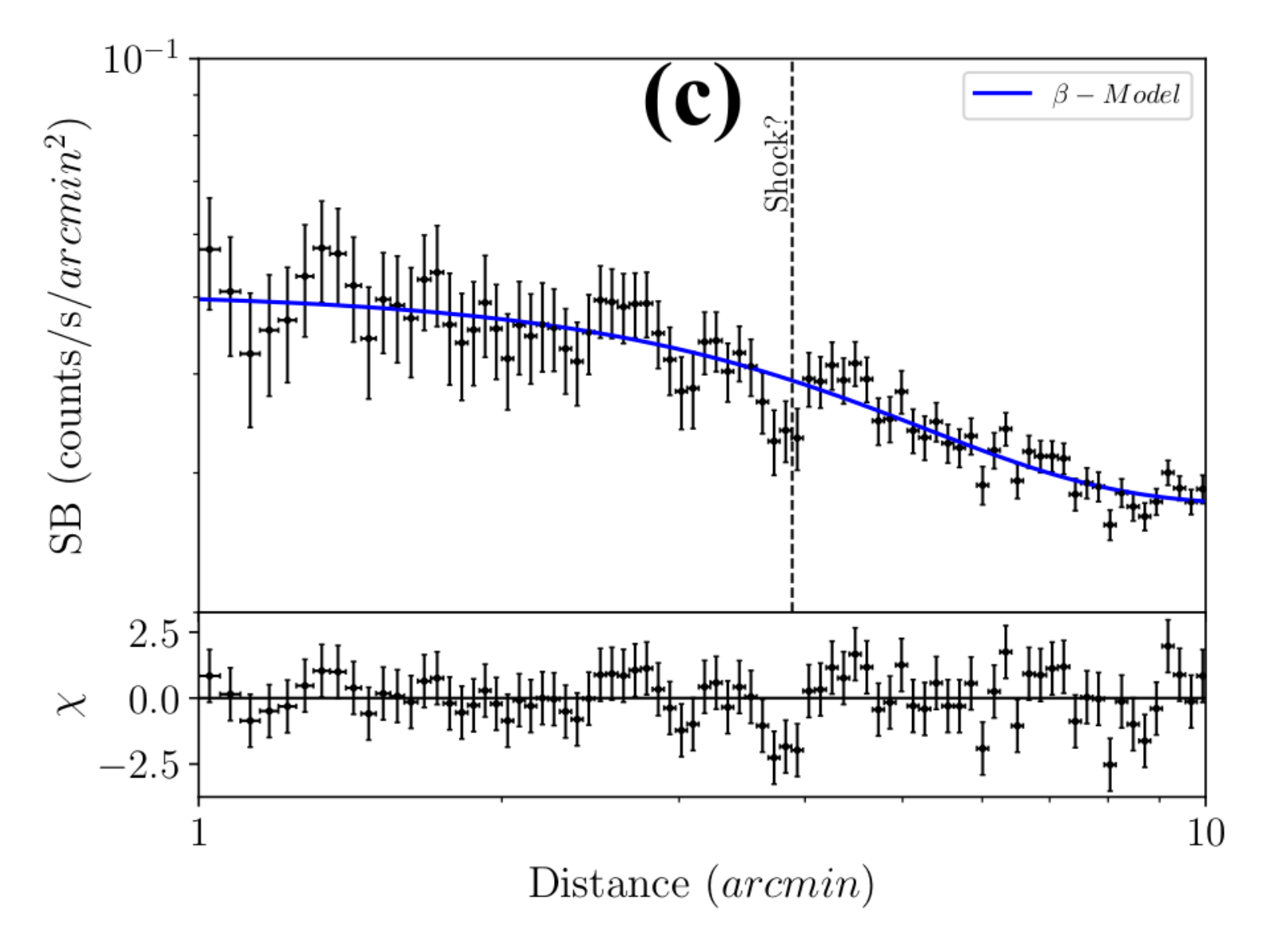}
\includegraphics[width=8.0cm]{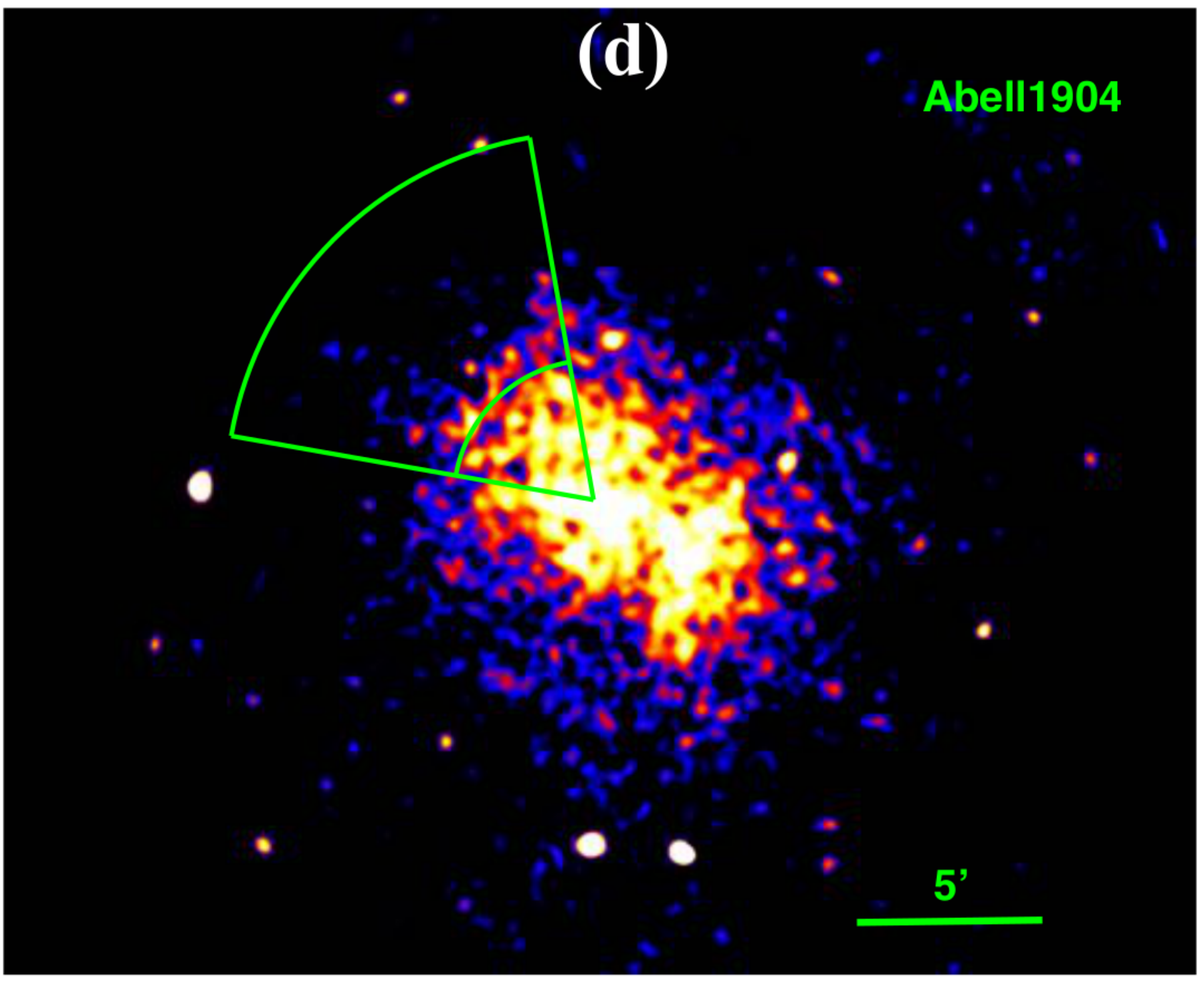}
\caption{{\bf Panel~(a):} Contours of uGMRT (Black, at$-3,3,5,7,9,18.. \times\sigma_{400}$) and LoTSS-I (Magenta, at $-3,3,6,9,18.. \times\sigma_{144}$) of PSZ2~G089.52+62.34 are plotted over the  PanSTARRS-I `$i$'-band image. {\bf Panel~(b):} X-ray contours plotted on the  uGMRT (grayscale) image and contour maps. The Planck~SZ centre is marked as blue `$\times$' and Abell cluster centre as green `+' for all the images in this paper. The physical scale for the images is 1.34 kpc/$\arcsec$. {\bf Panel~(c)}: The X-ray surface brightness profile fitted with $\beta-$model and its residues (for the sector $100^{\circ}-170^{\circ}$, as shown in {\bf Panel~(d)).}}\label{fig:radio-uGMRT-maps-psz089}
\end{figure*}

\subsection{A candidate radio relic in cluster PSZ2~G089.52+62.34}
%1

This is a low-redshift ($z$=$0.07008$) cluster with SZ mass $M_{500}^{sz}$= $1.83^{+0.19}_{-0.20}\times10^{14}{\rm{M_\odot}}$ (Planck-16). The rich galaxy cluster Abell~1904 is located well within this source, $\sim$4$\arcmin$ away from PSZ2~G089 centre. The reported 1D velocity dispersion of the Abell cluster is $803~{\rm{km}~s^{-1}}$ \citep{Struble_1999ApJS}, which is high value given its mass, and might be indicative of an ongoing merger. This is consistent with the elongated mass distribution along SW to NE \citep{Flin_2006A&A}, the evidence of substructures found in our XMM maps and the high X-ray luminosity 
($L_X=2.54^{+0.03}_{-0.02}\times10^{44}$~erg~s$^{-1}$) of the source.

A peripheral diffuse radio structure was noticed towards the NE of this cluster in LoTSS-I, low-resolution image (local $\sigma_{144}$=$200~{\rm{\mu Jy}~beam^{-1}}$; Fig.~\ref{fig:radio-uGMRT-maps-psz089}~(a)). In our uGMRT Band~3 image, we detect this peripheral radio emission as well, and it is shown as black contours within the blue-dashed polygonal area in Fig.~\ref{fig:radio-uGMRT-maps-psz089}(a). The largest linear size (LLS) and width of this elongated diffuse emission, as measured in the uGMRT maps, are 205~kpc and 45~kpc respectively. Its flux density is $S_{\rm{400}}=6.6\pm0.7$~mJy at uGMRT Band~3 and, for the same region, the LoTSS-I flux density is estimated to be $S_{\rm{144}}\sim19.9\pm3.1$~mJy. \citet{Weeren_2020arXiv} have reported a larger relic, as well as a second part of the relic, in their improved LoTSS-II images. This could not be detected in our  uGMRT maps (at a higher frequency) at the reported RMS noise level of $50~\mu$Jy~$\rm{beam}^{-1}$. Such morphological dissimilarity between images at different frequencies is common for steep-spectrum radio sources.

In Fig.~\ref{fig:radio-uGMRT-maps-psz089}~(a), uGMRT contours are overlaid on  a PanSTARRS~1 optical image, where no optical counterpart of this diffuse radio emission is seen. XMM X-ray contours show two prominent substructures, marked as `A' \& `B' in Fig.~\ref{fig:radio-uGMRT-maps-psz089}~(b), possibly in the process of merger along the SW to NE direction. As is usual for radio relics, the convex and elongated radio structure at the outskirts of this cluster coincides with the outer X-ray contours, placed well ahead of the merging clumps, perpendicular to the merging axis (see  Fig.~\ref{fig:radio-uGMRT-maps-psz089}~(b)). This structure is found about $4\arcmin$ away from the bigger X-ray substructure (`A'). Though clear evidence of a thermal shock is absent in the low-exposure (15~ks) XMM data, interestingly, a sudden dip in the X-ray surface brightness profile, for the sector $100^{\circ}-170^{\circ}$ shown in Fig.~\ref{fig:radio-uGMRT-maps-psz089}~(d)), with a residual of more than $2\sigma$ from the $\beta$-model fit (see Fig.~\ref{fig:radio-uGMRT-maps-psz089}~(c)), certainly indicates the presence of an apparent shock front at a similar distance ($\sim4\arcmin$) as the radio diffuse emission. 

Since \citet{Weeren_2020arXiv} have provided the combined flux density for two parts of the relic and we could detect only one part of it in our uGMRT maps, it was not possible to compute a meaningful spectral index. We have however made a rough estimate of an average spectral index between the uGMRT detection and its co-located LoTSS-I counterpart,  found to be $\alpha_{144}^{400} \sim-1.08$, suggesting a case of relic emission. In the absence of stronger evidence, we classify it as a candidate relic.

\subsection{A radio relic and trailing emission in PSZ2~G111.75+70.37}
%2

\begin{figure*}
\includegraphics[width=8.5cm]{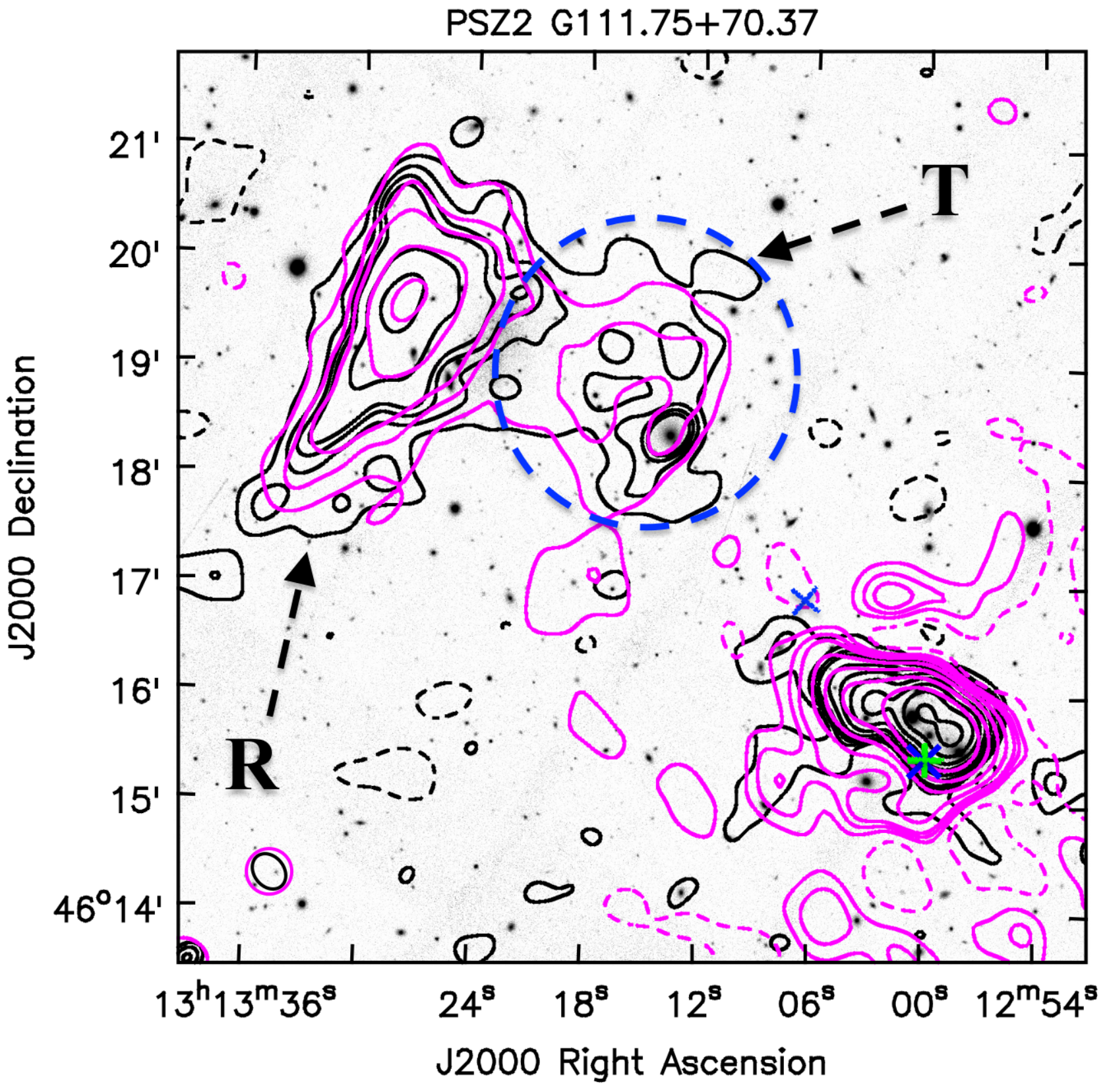}
\includegraphics[width=8.5cm]{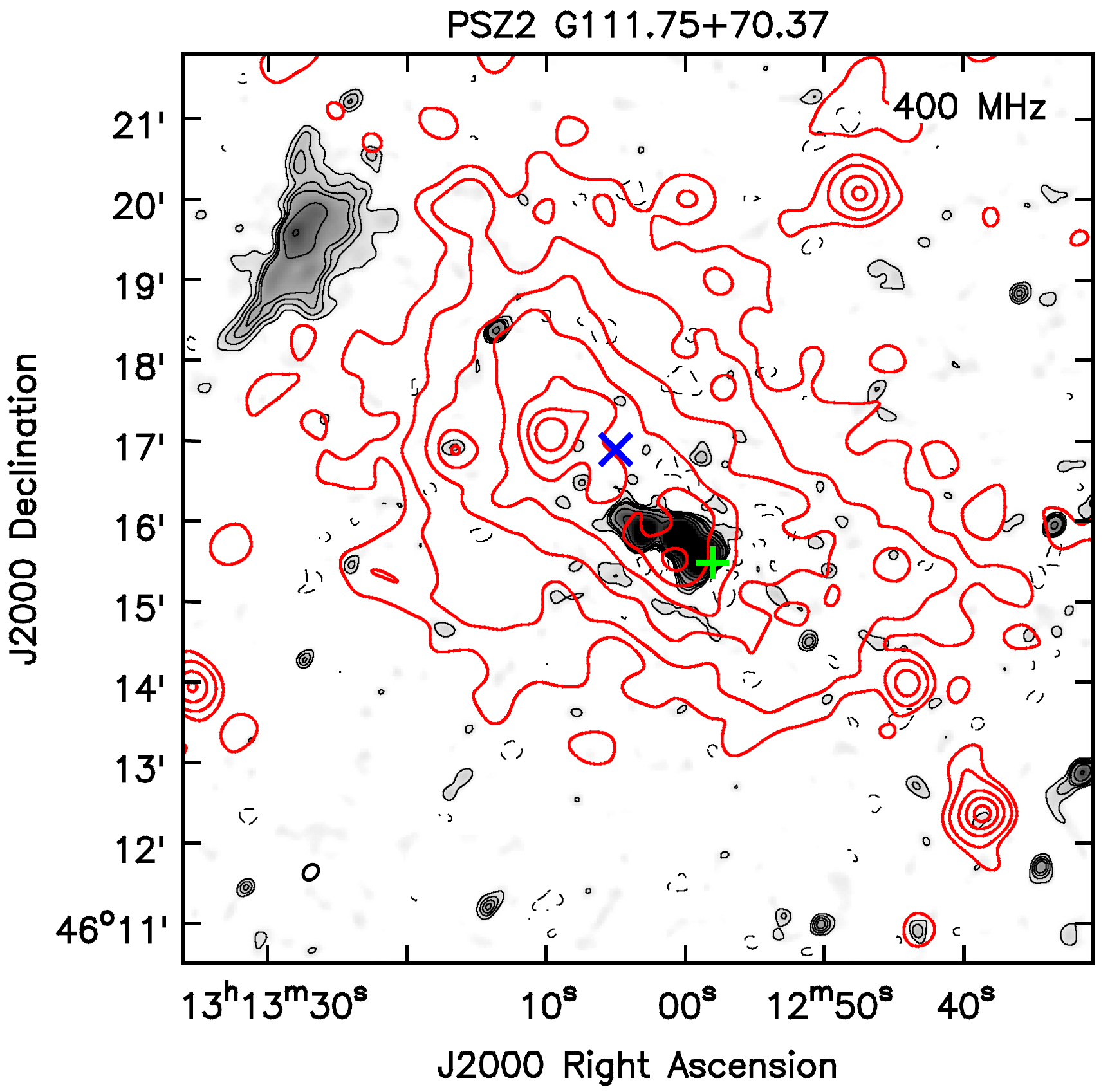}\hspace{-0.1cm}
\caption{{\bf Panel~(a):} uGMRT and LoTSS-I contours are plotted on PanSTARSS-I `$i$'-band image of source PSZ2~G111.75+70.37. Contours (levels same as Fig.~\ref{fig:radio-uGMRT-maps-psz089}) of uGMRT (Black), LoTSS-I (Magenta) with RMS noise of 100 and 420 $\mu$Jy~beam$^{-1}$, respectively. {\bf Panel~(b):} XMM contours (red) are plotted over uGMRT high resolution grayscale image and contours (at -3,3,6,9,18,36$\times\sigma$, where $\sigma=65~\mu$Jy~beam$^{-1}$) maps. The physical scale for the images is 3.05~kpc/$\arcsec$. }\label{fig:radio-uGMRT-maps-psz111}
\end{figure*}

\begin{figure}
\includegraphics[width=8.5cm]{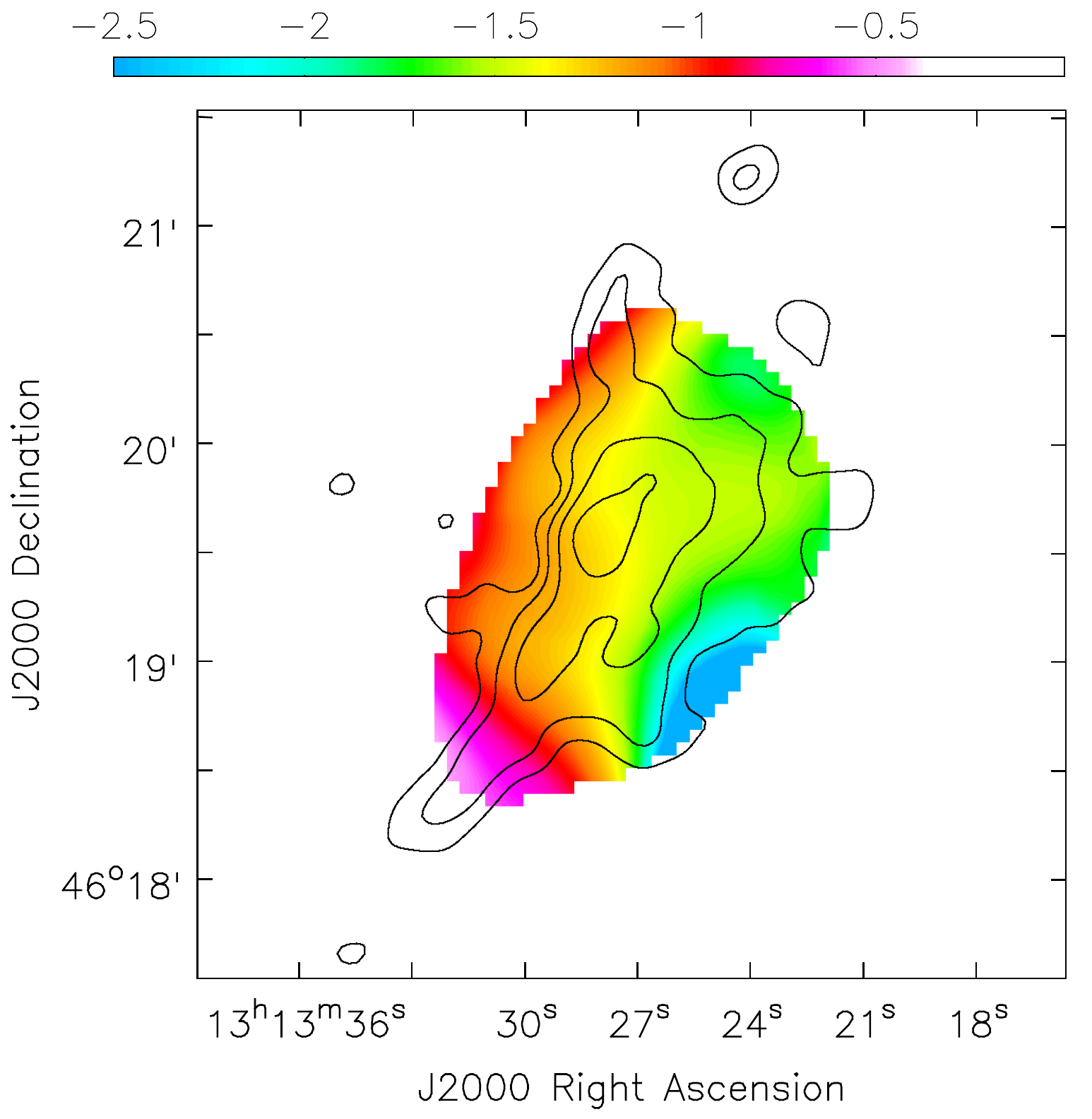}
\caption{Spectral index map of the relic found in the cluster PSZ2~G111.75+70.37, prepared using the uGMRT (400~MHz) and NVSS (1.4~GHz) images}.\label{fig:SI-map-uGMRT-NVSS}
\end{figure}

PSZ2~G111.75+70.37, identified with the poor cluster Abell~1697 is a moderately distant cluster at a spectroscopic redshift of $z=0.1813$. The estimated mass for the cluster in the Planck-16 catalogue is $M^{SZ}_{500}=4.34^{+0.32}_{-0.33}\;\times10^{14}\;\rm{M_{\odot}}$  and its estimated XMM X-ray luminosity is $L_{X}=5.26^{+0.30}_{-0.40}\times10^{44} \rm{erg\;s^{-1}}$. This cluster is reported to host a radio relic and a ultra-steep spectrum trailing emission, as tentatively estimated from LoTSS-I and NVSS maps \citep{Paul_2020A&A}.

In this case, we detect both the radio structures in our uGMRT band~3 low-resolution map (see Fig.~\ref{fig:radio-uGMRT-maps-psz111}~(a); RMS noise $\sigma_{400}=100~\mu$Jy~beam$^{-1}$), similar to those seen in LoTSS-I ($\sigma_{144}$=$420~{\rm{\mu Jy}~beam^{-1}}$). However, no clear evidence of radio halo emission, as found by \citet{Weeren_2020arXiv},
could be seen in our maps.

The relic is 750~kpc long and 350~kpc in width (marked as `R' in Fig.~\ref{fig:radio-uGMRT-maps-psz111}~(a)), having a uGMRT flux $S_{400}(R)=49.7\pm5.0$~mJy. The distance from the Planck cluster centre to the peak relic emission is about 850~kpc. The trailing emission (marked as `T' and enclosed within the dashed ellipse) is large, about $450\times350$ kpc in projected extent, but is very faint ( $S_{400}(T)=15.3\pm1.6$~mJy). 

The LoTSS-I flux for the two detected structures and the region same as uGMRT are $S_{144}(R)\sim130.2\pm26.1$~mJy and $S_{144}(T)\sim41.7\pm8.3$~mJy. Accurate spectral index estimation with these values would not be possible because of the fact that the LoTSS-I map that is available to us for this object is contaminated with significant artefacts as mentioned in \citet{Weeren_2020arXiv}. 

Fortunately, the NVSS map for this cluster contains a detection of a major part of the relic. Combining the images at 1.4 GHz (NVSS) and the matched $uv$-coverage map at 400 MHz (uGMRT), we produced a spectral index map as shown in Fig.~\ref{fig:SI-map-uGMRT-NVSS}. The average spectral index of the proposed relic part (i.e. ‘R’, common area with $3\sigma$ detection
of NVSS) is about $\alpha_{int}{^{1420}_{400}}=-1.40\pm0.01$. The spectral index at the relic front is $\alpha_{inj}=-0.93\pm 0.01$, which gradually steepens to $\alpha_{tail}=-1.87\pm0.01$ towards the inside edge of the relic. Therefore, it reasonably follows the predictions of the commonly-used continuous injection spectrum model for radio relics i.e., $\alpha_{inj}{^{1420}_{400}}=-0.93\sim \alpha_{int}{^{1420}_{400}}+0.5$.

Our XMM-Newton contours in Fig.~\ref{fig:radio-uGMRT-maps-psz111}~(b) clearly show substructures merging along the SW to NE direction, and the relic is found ahead of this at the cluster outskirts, perpendicular to the merging axis. This is suggestive of a typical binary merger of clusters producing peripheral radio relics. All the possible scenarios for the origin of radio emission in this cluster have been discussed in detail in \citet{Paul_2020A&A}.

\subsection{A radio halo-relic system in cluster PSZ2~G080.16+57.65}
%3

\begin{figure*}
\includegraphics[width=8.5cm]{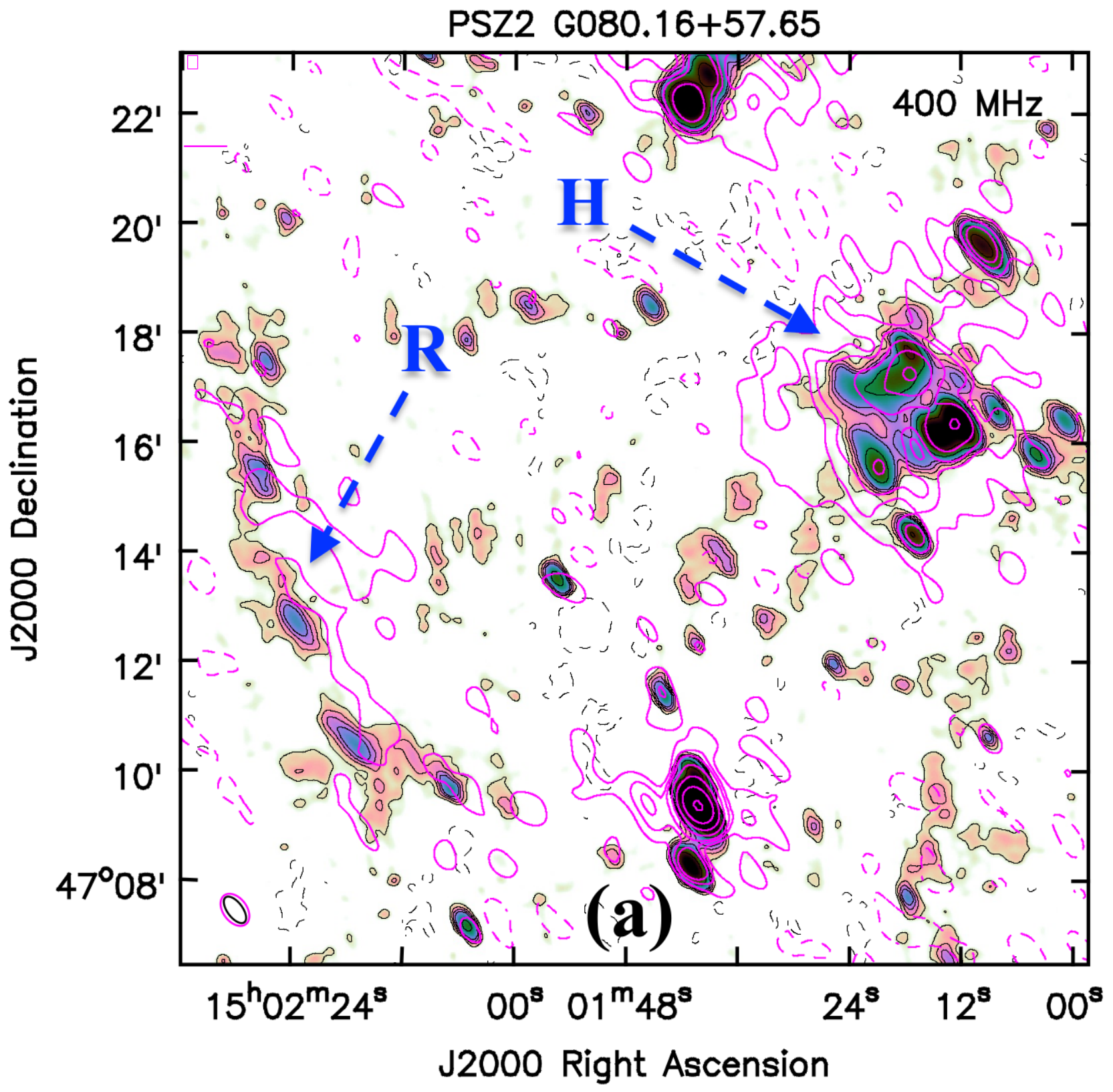}\hspace{-0.1cm}
\includegraphics[width=8.5cm]{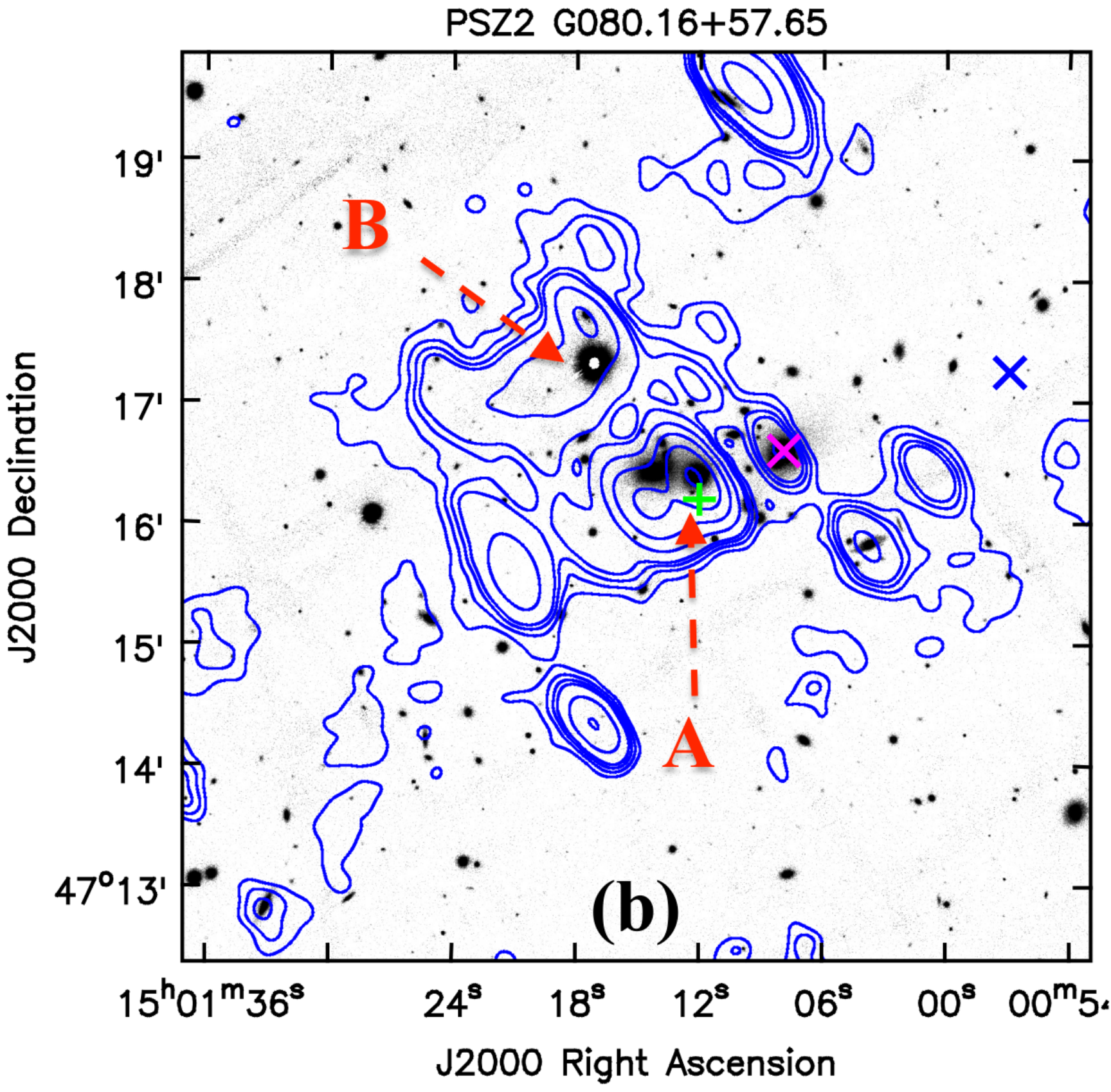}\hspace{-0.1cm}\\
\caption{{\bf Panel~(a):} uGMRT high-resolution colour map and contours (black) and LoTSS-I contours (magenta) for the cluster PSZ2~G080.16+57.65 (contours are at the same level as Fig.~\ref{fig:radio-uGMRT-maps-psz089}~(a) with RMS noise $\sigma=55$ and $240~\mu$Jy~beam$^{-1}$, respectively). {\bf Panel~(b):} uGMRT contours (blue) plotted over the PanSTARRS-I `$i$'-band image, zoomed into the central halo region. The centre of the GMBCG J225.28318+47.27663 cluster, clearly associated with this source, is shown as a magenta cross. The physical scale for the images is 1.64 kpc/$\arcsec$.} \label{fig:radio-uGMRT-maps-psz080} 
%\vspace{-0.3cm}
\end{figure*}

\begin{figure}
\includegraphics[width=8.5cm]{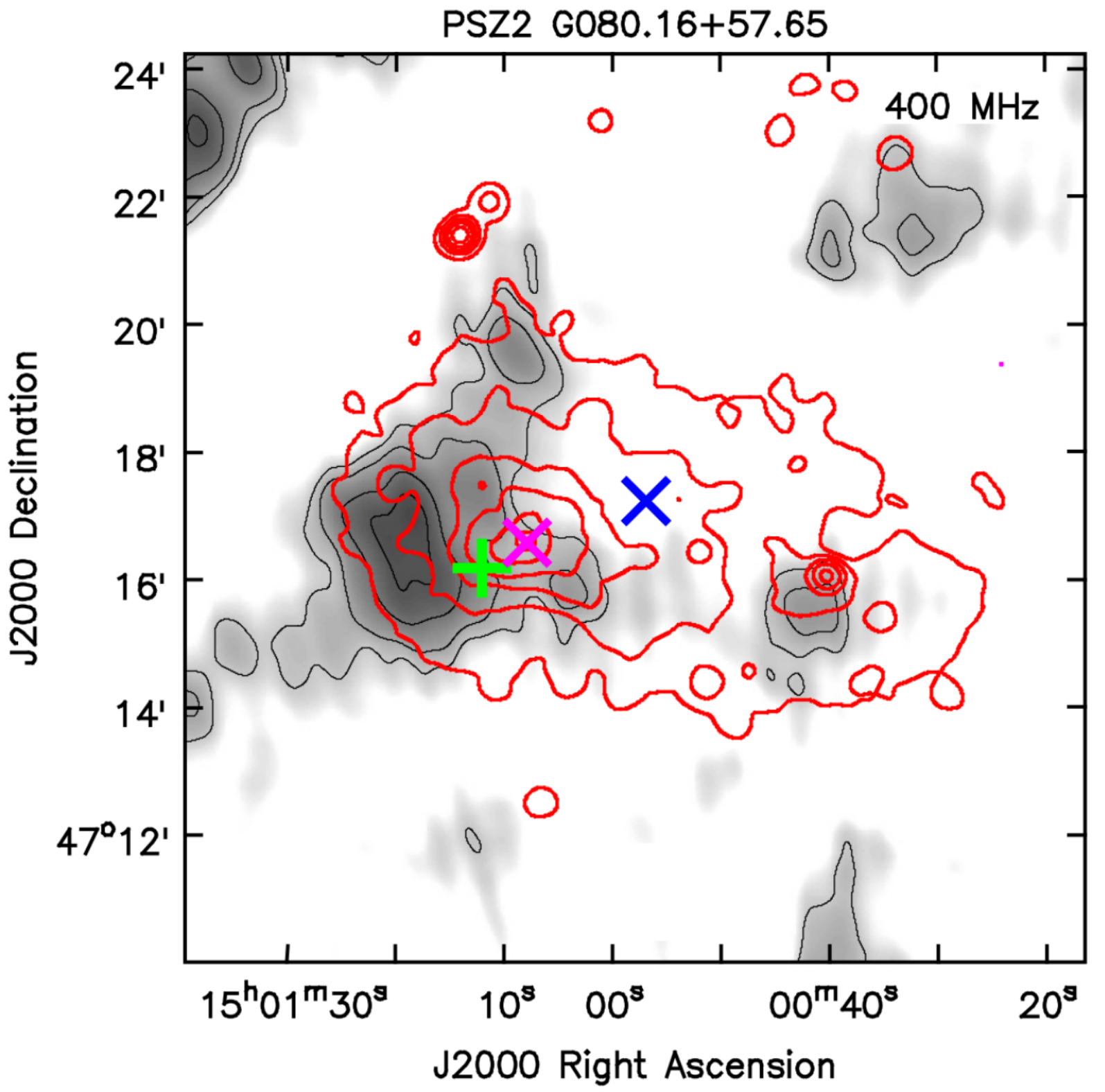}\hspace{-0.1cm}
\caption{XMM contours (red) over uGMRT point source subtracted map of the source PSZ2~G080.16+57.65.} \label{fig:radio-uGMRT-Xray-psz080} 
%\vspace{-0.3cm}
\end{figure}

PSZ2~G080.16+57.65 (or PSZ2~G080 in short), also identified with the cluster Abell~2018, is another low redshift ($z=0.0878$) poor cluster in our sample, having mass $M_{500}^{sz}=2.51^{+0.20}_{-0.21}\times10^{14}{\rm{M_\odot}}$ (Planck-16), and a moderate X-ray luminosity $L_{\rm{X}}=8.14^{+0.04}_{-0.03}\times10^{43}$~erg~s$^{-1}$. A cluster named GMBCG~J225.28318+47.27663 \citep{Hao_2010ApJS} shown as a magenta cross in Fig.~\ref{fig:radio-uGMRT-maps-psz080}~(b), is within $2\arcmin$ of PSZ2~G080, with a comparable redshift $z=0.0882$ and mass $M=2.86\pm0.26\times10^{14}M_\odot$, indicating that they are the same object. 

We found a large, 1~Mpc peripheral bow-like structure and an indication of diffuse radio emission in the central region of this cluster in our uGMRT high-resolution map (see Fig.~\ref{fig:radio-uGMRT-maps-psz080}~(a)). The peripheral structure in this cluster is very faint with a flux $S_{400}=13.3\pm1.4$~mJy, but at an unusual distance of 1.3~Mpc away from the cluster centre, beyond the virial radius (self-similar fiducial $r_{200}\sim1.2$~Mpc). The spectral index of $\alpha{^{400}_{144}}\sim-1.4$, as computed using our uGMRT and 144~MHz (reported in \citet{Weeren_2020arXiv}) fluxes, as well as, its peripheral location and bow-like shape suggest it to be a steep-spectrum radio relic.

The only radio point source found inside this cluster (in NED) is a radio galaxy, exactly at the centre of the Abell cluster (green `+' in Fig.~\ref{fig:radio-uGMRT-maps-psz080}~(b)), and at a similar redshift $z=0.0873$, confirming it to be a member of the cluster. This also matches with a radio peak seen in our uGMRT map (see `A' in Fig.~\ref{fig:radio-uGMRT-maps-psz080}~(b)). The other bright but extended radio source (`B') in uGMRT and LoTSS-I maps cannot be seen in VLA First or any other higher-frequency radio surveys. There is no optical source in NED identified as a galaxy that coincides with `B', which  confirms its nature as a diffuse radio source. 

We produced a point-source subtracted image to map the central diffuse emission in this cluster, following the method described in \S~\ref{sec:data}. We found a halo-like emission with a size $\sim540\times490$~kpc and flux density  $S_{400}=19.3\pm2.0$~mJy (see Fig.~\ref{fig:radio-uGMRT-Xray-psz080}) within $3\sigma$ contours. Unlike \citet{Weeren_2020arXiv}, we found no radio bridge connecting the relic and the halo, in our uGMRT point-source-removed maps, which have an RMS of $\sigma=120~\mu$Jy~beam$^{-1}$.

Fig.~\ref{fig:radio-uGMRT-Xray-psz080} reveals an elongated and clumpy X-ray map, indicative of a non-relaxed state. The radio structure uncovered here is found to be well-enclosed by the bigger X-ray clump, the peak of which also coincides with the centre of cluster GMBCG~J225 (magenta `$\times$'), establishing its connection to the ICM of the cluster. Following an alternative method of estimating the flux density of the  radio-halo, as described in \S~\ref{sec:data}, we computed a flux density of $18.0\pm0.01$~mJy within the 3~e-folding radius (see Tab.~\ref{tab:findings}). In comparison, \citet{Weeren_2020arXiv} computed a flux density of 92~mJy for the same halo, giving rise to an average spectral index $\alpha{^{400}_{144}}\sim-1.6$. Though this is indicative of an ultra-steep spectrum radio halo, the radio emission is neither centrally located as in archetypal radio haloes, nor found at the cluster outskirts as are the radio relics, making it difficult to firmly assign an identity to it. 

\subsection{PSZ2~G106.61+66.71, an intermediate halo?}
%4

\begin{figure*}
\includegraphics[width=8.5cm]{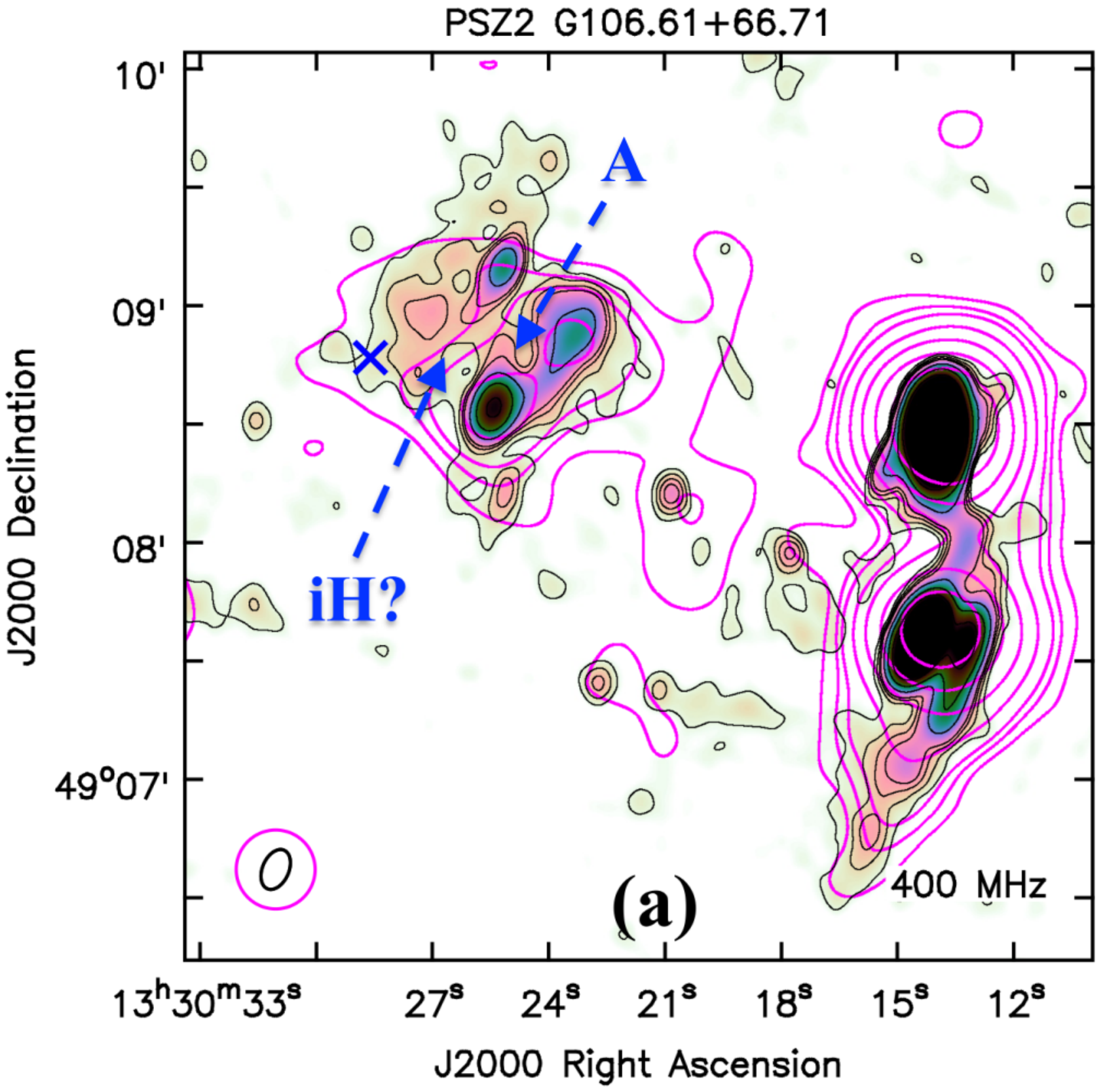}
\includegraphics[width=8.5cm]{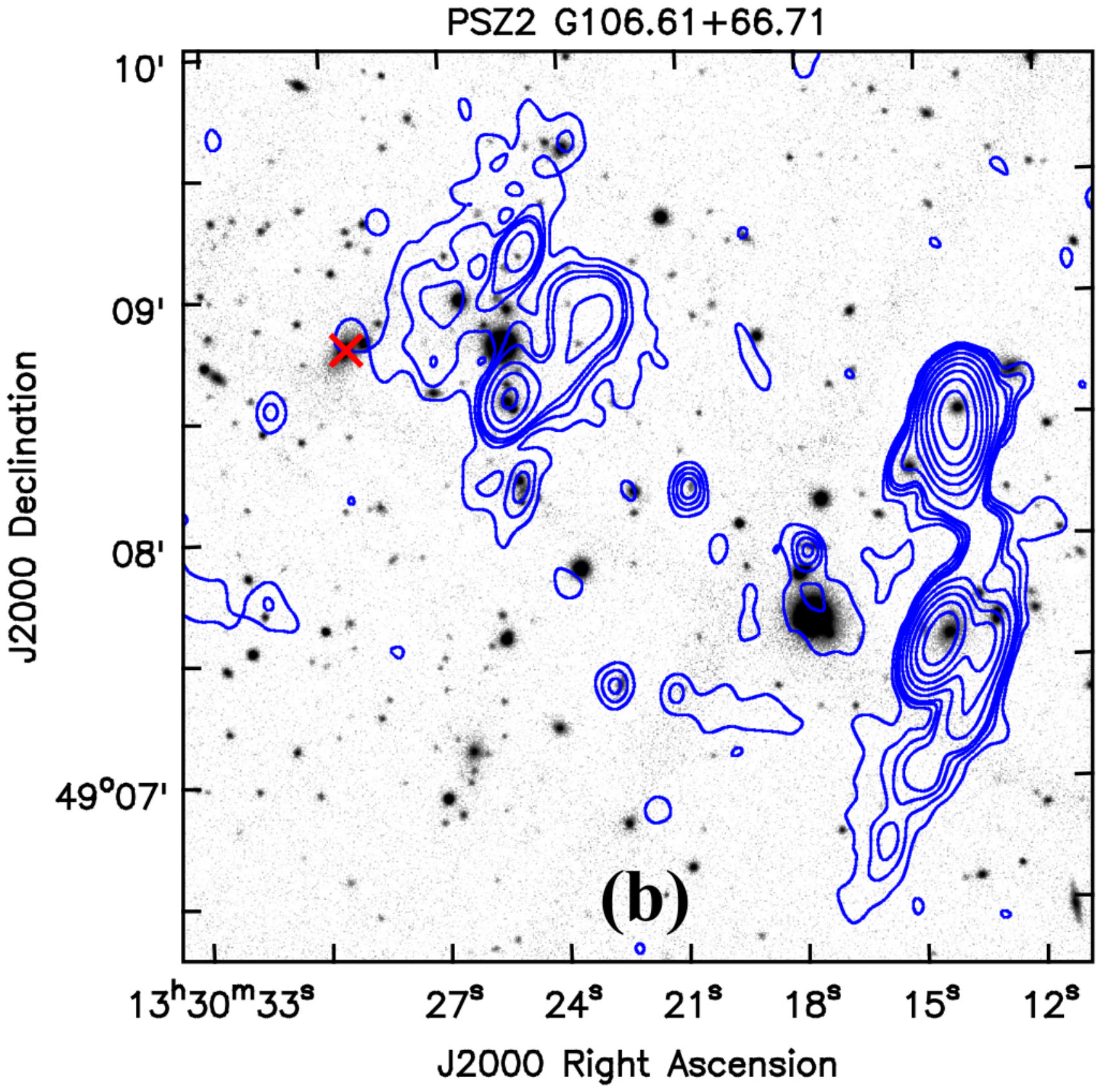}\\
\includegraphics[width=8.5cm]{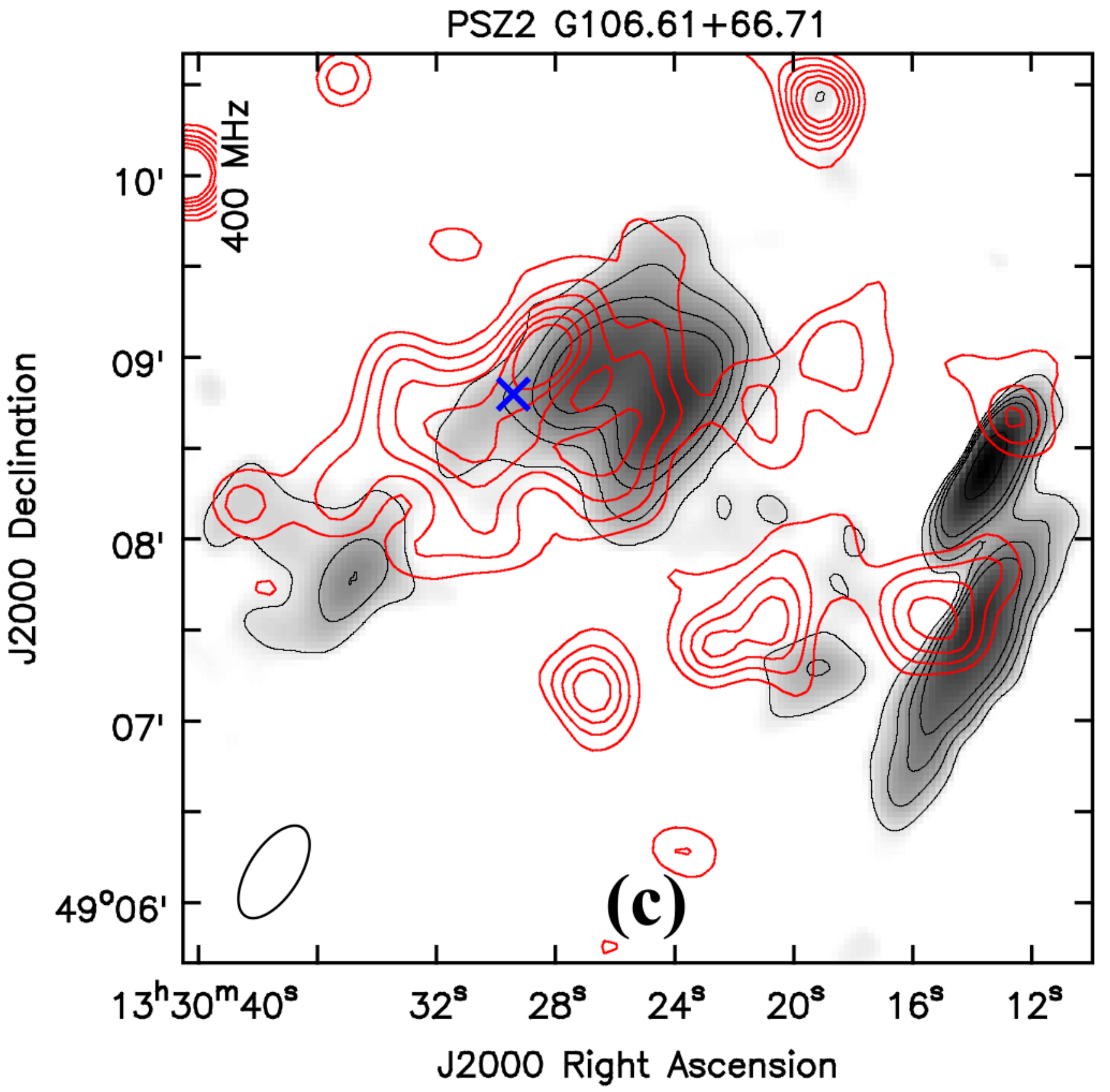}
\includegraphics[width=8.9cm]{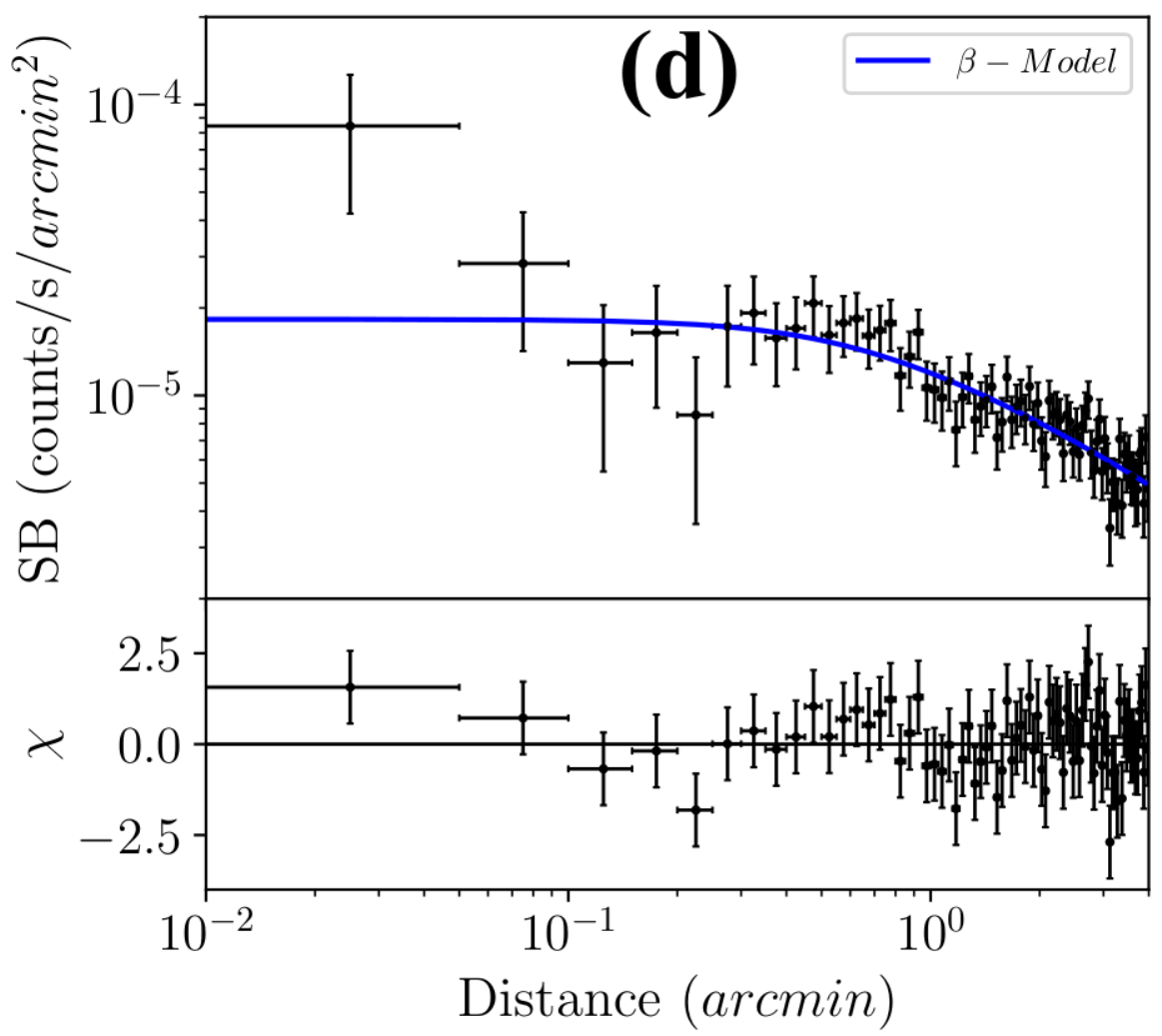}
\caption{{\bf Panel~(a):} uGMRT high-resolution colour map and contours (black) and LoTSS-I contours (magenta) for the cluster PSZ2~G106.61+66.71 (contours are at the same level as Fig.~\ref{fig:radio-uGMRT-maps-psz089}~(a) with RMS noise $\sigma=45$ and $300~\mu$Jy~beam$^{-1}$, respectively). {\bf Panel~(b):} uGMRT contours (blue) over PanSTARRS-I `i'-band image. {\bf Panel~(c):} {\it Chandra} contours (red) over uGMRT point source subtracted map. Radio contours are at $3,5,7,9,18,36..\times\sigma$ with $\sigma=120~\mu$Jy~beam$^{-1}$. {\bf Panel~(d)}: X-ray surface brightness profile fitted with $\beta-$model and the residuals for the circular annuli centred at the X-ray peak. The physical scale for the images is 4.76 kpc/$\arcsec$.} \label{fig:radio-uGMRT-maps-psz106} 
\end{figure*}

PSZ2~G106.61+66.71 is the highest redshift ($z=0.33140$) cluster in our sample detected with diffuse radio emission. This rich cluster with 82 galaxy members within a radius $r_{200}=1.76$~Mpc \citep{Wen_2013MNRAS} with an SZ mass $M_{500}^{sz}=4.67^{+0.55}_{-0.57}\times10^{14}{\rm{M_\odot}}$ (Planck-16). The BCG in this cluster is a radio bright source (NVSS~J133024+490846, `A') and about 180 kpc away from the {\it Chandra} X-ray peak (see Fig.~\ref{fig:radio-uGMRT-maps-psz106}~(a)\&(c)).

Both uGMRT and LoTSS-I ($\sigma_{144}$=$300~{\rm{\mu Jy}~beam^{-1}}$) radio maps show the indication of an extended diffuse emission around the BCG (see Fig.~\ref{fig:radio-uGMRT-maps-psz106}~(a)). To study the diffuse emission in detail, we produced a point source subtracted uGMRT image by the method described in \S~\ref{sec:data}. The contours of this image (black; Fig.~\ref{fig:radio-uGMRT-maps-psz106}~(c)) reveals that the diffuse emission has an extent of at least $\sim550\times530$~kpc \st{(at $5\sigma$)}. 
% flux density $S_{400}=8.6\pm1.0$~mJy
The computed flux density  (using the profile fit method described in \S~\ref{sec:data}) is $6.6\pm 0.01$~mJy within the 3~e-folding radius i.e., $3r_e=507\pm17$ (see Tab.~\ref{tab:findings}) for our uGMRT  map. At 144~MHz,  using the same method, the flux is estimated to be 20~mJy \citep{Weeren_2020arXiv}. This leads to  an average spectral index of $\alpha{^{400}_{144}}\sim-1.09$.

The value of the radio power estimated from these values is rather high $P_{\rm{1.4GHz}}=6.34\pm0.19\times10^{23}~{\rm{W~Hz^{-1}}}$ at 1.4~GHz. Such an emission feature is usually seen in cool-core clusters as radio mini-haloes, albeit usually with an extent less than 500~kpc. Our {\it Chandra} X-ray analysis shows a low-brightness core with non-smooth profile (see Fig.~\ref{fig:radio-uGMRT-maps-psz106}~(d)), usually a feature of a weak cool-core cluster. X-ray maps also show elongation and substructures (red contours in Fig.~\ref{fig:radio-uGMRT-maps-psz106}~(c)). The total  X-ray luminosity of the cluster is on the high side $L_{\rm{X}}=3.97^{+0.4}_{-0.3}\times10^{44}$~erg$s^{-1}$, given its velocity dispersion, indicative of a merger. 

This is not the typical scenario for a mini-halo found in relaxed, cool-core clusters. Thus, a second possibility is that the cool-core in this cluster has been dislodged from its original location due to a recent merger. Radio emission in PSZ2~G106, therefore, may be an intermediate radio halo, similar to other cases reported by, for instance, \citet{Savini_2018MNRAS,Kale_2019MNRAS} and \citet{Raja_2020MNRAS}.

\subsection{Correlations between radio power of haloes and relics and other parameters}\label{correlation}

\begin{table}
\caption{Parameters for $P_{1.4\ {\rm GHz}}$--$M_{500}$ and $P_{1.4\ {\rm GHz}}$--LLS correlations}
\centering
\small
\begin{tabular}{rccc}
\hline
Object name & Spectral index & $P_{1.4\ {\rm GHz}}$ & Relic LLS \\ 
 (Structure) & & $(10^{23}{\rm{W~Hz^{-1}}}$)& (kpc) \\
\hline
PSZ2~G080~(H) & -1.60 & $0.49 \pm 0.01$ & -- \\
(R) & -1.40 & $0.46 \pm 0.01$ & 1000 \\
PSZ2~G089~(R)& -1.08 & $0.20 \pm 0.05$ & 205 \\
PSZ2~G106~(H)& -1.08 & $6.34 \pm 0.19$ & -- \\
PSZ2~G111~(R)& -1.40 & $8.68 \pm 1.26$& 750 \\
\hline
\end{tabular}\label{tab:flux-power-size}
%\vspace{-0.3cm}
\end{table}

\begin{figure}
\includegraphics[width=8.8cm]{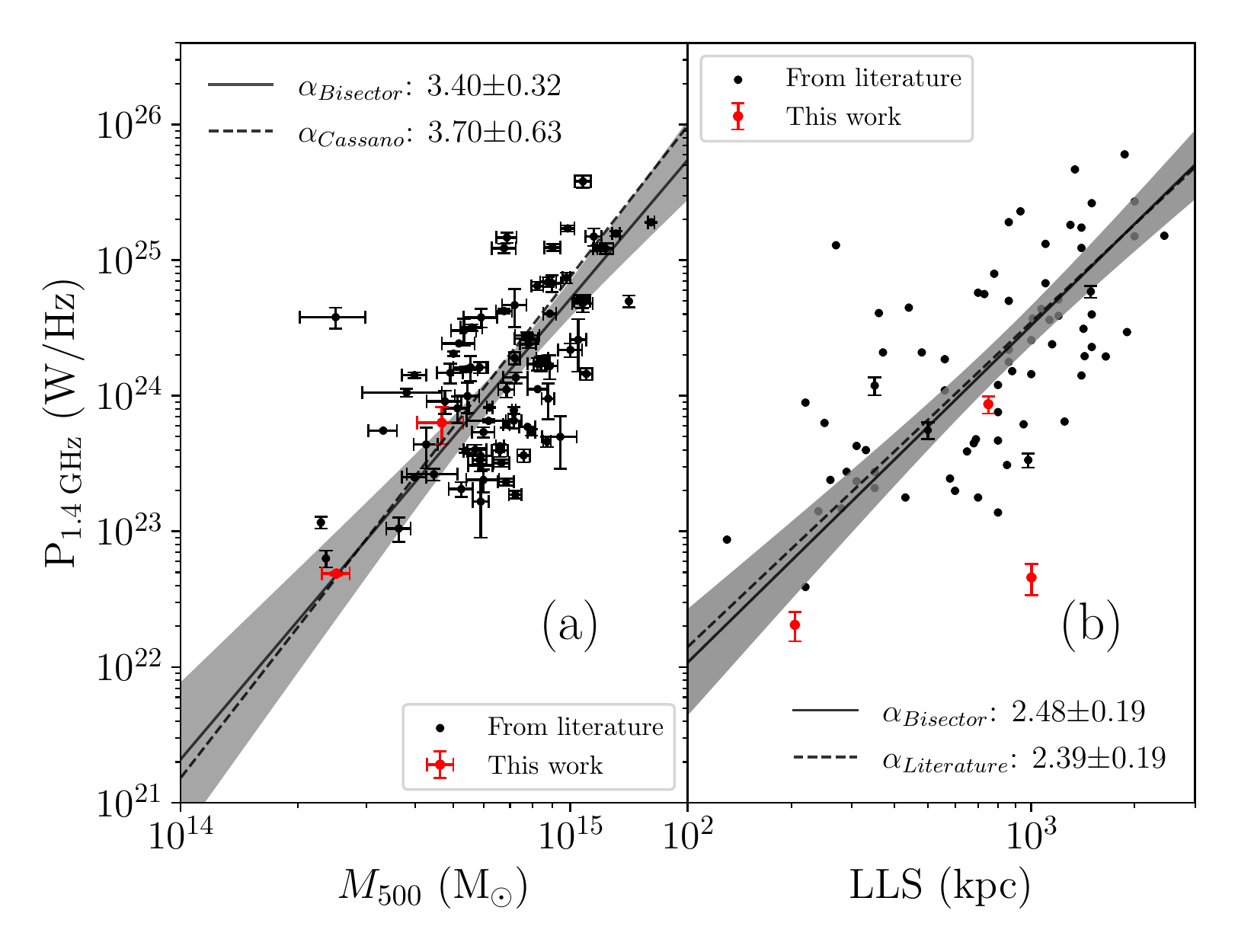}
\caption{{\bf Panel~(a):} The radio power at 1.4 GHz  of radio haloes vs the mass of the host cluster within $R_{500}$ ($M_{500}$) and {\bf Panel~(b):} The radio power at 1.4 GHz  of radio relics ($P_{1.4\ {\rm GHz}}$)
vs the largest linear size (LLS) of the relics. The objects studied in this paper are marked in red. The samples are discussed in \S~\ref{correlation}.}\label{fig:correlation-plot}
\end{figure}

Following the early work by \citet{Venturi_2008A&A}, which helped to establish the link between the existence of radio haloes and the recent merger history of clusters, the correlation of halo power (commonly estimated at 1.4~GHz, e.g. \citet{Cassano_2010A&A}) with various cluster parameters, gives insights into the origin and nature of these haloes, and helps us further establish the physical processes responsible for these systems. 

In this context, the widely used correlation for cluster mass ($M_{500}$) and radio halo power at 1.4~GHz ($P_{1.4}$) has been computed by \citet{Cassano_2013ApJ} using a sample of predominantly massive clusters ($>5\times10^{14}{\rm{M}}_{\sun}$). A correlation has also been reported for the relic power (at 1.4~GHz) and its size (LLS) by \cite{Feretti_2012A&ARv} and recently updated by \citet{Paul_2020A&A}. However, as discussed and reported in this study, in the age of uGMRT and LoFAR, as the number of detections dramatically increase, one needs to validate the scaling relations, especially as detections are pushed to lower cluster masses.

In this paper, in all, we detect two possible radio haloes and three radio relics. In order to include them in this  correlation study, the observed radio power with uGMRT at 400 MHz is scaled to that at 1.4~GHz, assuming the spectral indices reported in this paper (see Tab.~\ref{tab:flux-power-size}).  SZ masses ($M{^{SZ}_{500}}$) are used for all the clusters for which they are available.

Fig.~\ref{fig:correlation-plot}~(a), the correlation plot of radio halo power ($P_{\rm{1.4GHz}}$) vs the cluster mass ($M_{500}$) consists the data from \citet{Cassano_2013ApJ}, updates from \citet{Weeren_2019SSRv} and two objects reported in this paper. While, the \citet{Cassano_2013ApJ} sample yields a BCES bisector slope $\alpha_{BCES}=3.70\pm0.63$, the combined correlation slope with all the above mentioned data is found to be $\alpha_{comb}=3.40\pm0.32$, a slightly flatter, though having well overlapping error limits. 

The data for radio relic power ($P_{\rm{1.4GHz}}$) vs the relic size ($LLS$) correlation plot shown in Fig.~\ref{fig:correlation-plot}~(b) is largely updated from  \cite{Weeren_2019SSRv,Paul_2019MNRAS} and three from this study. A new slope $\alpha_{new}=2.48\pm0.19$ is found to be fairly consistent with the correlation of $\alpha=2.39\pm0.19$ computed from data found in the literature. We note that the relic of PSZ2G~080 is a far outlier, having a very low radio emission power considering its large LLS. Interestingly, this is so far the only relic found beyond the virial radius ($>r_{200}$) of the host cluster.

\vspace{-0.5cm}
\section{Summary and conclusions}\label{sec:conc} 

We present the results from the first systematic search for diffuse radio emission from low-mass clusters ($<5\times10^{14}{\rm{M_\odot}}$), in the overlap region of  the Planck~SZ catalogue (PSZ2) and the LoTSS data release~1, confirmed by deep uGMRT maps at 400~MHz obtained by us. Out of 12 clusters in our sample (see \S~\ref{sec:data}), four are detected with diffuse radio emission. The peripheral structures detected in PSZ2~G111 and PSZ2~G080 clusters are the steep-spectrum relics. In the absence of accurate spectral properties, similar emission found in PSZ2~G089 is currently speculated to be that from a relic. Cluster PSZ2~G080 also hosts a radio halo along with the relic, and PSZ2~G106 is detected only with a central diffuse emission, argued to be an intermediate radio halo. 

In this context, it is important to mention that the cluster PSZ2~G089, with $M^{SZ}_{500}=1.8\times10^{14}$~M$_\odot$, is among the lowest mass clusters detected with a radio relic, and PSZ2~G080 with $M^{SZ}_{500}=2.5\times10^{14}$~M$_\odot$ is the lowest mass cluster to host both a radio halo and a Mpc size relic. In the latter case, the relic, at a distance 1.3 Mpc  from the cluster centre, is the only such specimen in a cluster with the relic found beyond the virial radius, and is an outlier in the scaling relation between radio relic power and spatial extent. With these new detections, we update the $P_{1.4\ {\rm GHz}}$--$M_{500}$ correlation in a much wider range of masses, especially towards the lower mass end.

We note that the unavailability of corrected LoTSS-I images (i.e. LoTSS-II images) in the public domain restricted us from making the spectral index maps that would provide further indications for the accurate classification of these sources. Nevertheless, the results obtained from our sample consisting of a small number of low-mass clusters are crucial, considering their rarity, and the implications of their existence on the study of non-thermal emission mechanisms over a wide range of masses of such systems. This also motivates us to examine low-mass clusters with SKA precursors such as uGMRT and MeerKAT etc. with a larger data set.

\section*{Acknowledgements}
This research was funded by DST INSPIRE Faculty Scheme awarded to SP (code: IF-12/PH-44). PG acknowledges Council of Scientific \& Industrial Research (CSIR) for a Senior research fellowship (CSIR-SRF; file no: 09/137(0574)/2018 EMR-1). S. Salunkhe wants to thank "Bhartratna JRD Tata Gunwant Sanshodhak Shishyavruti Yojna" for a doctoral fellowship. S. Sonkamble acknowledges financial contribution from ASI-INAF n.2017-14-H.0 (PI A. Moretti). We thank the Director, NCRA-TIFR, for granting discretionary time on uGMRT, and the staff of the GMRT for making these observations possible. GMRT is run by the National Centre for Radio Astrophysics of the Tata Institute of Fundamental Research. \\

\noindent{\bf Data availability:} \\
1. uGMRT data can be accessed from the GMRT data archive: https://naps.ncra.tifr.res.in/goa/data/search. The calibrated data can be shared on reasonable request to the corresponding author.\\
\noindent 2. LoTSS-I data sourced from public domain archive: https://lofar-surveys.org/surveys.html.
\vspace{-0.7cm}



\bibliographystyle{mnras}

%\begin{thebibliography}{99}
%\end{thebibliography}

%%%%%%%%%%%%%%%%%%%%%%%%%%%%%%%%%%%%%%%%%%%%%%%%%%

%%%%%%%%%%%%%%%%% APPENDICES %%%%%%%%%%%%%%%%%%%%%

%\appendix

%%%%%%%%%%%%%%%%%%%%%%%%%%%%%%%%%%%%%%%%%%%%%%%%%%

% Don't change these lines
\bsp	% typesetting comment
\label{lastpage}
\end{document}